# How to evaluate universities in terms of their relative citation impacts: Fractional counting of citations and the normalization of differences among disciplines

Loet Leydesdorff [1] & Jung C. Shin [2]

(alphabetical order; no seniority implied)

**Abstract**

Fractional counting of citations can improve on ranking of multi-disciplinary research units (such as universities) by normalizing the differences among fields of science in terms of differences in citation behavior. Furthermore, normalization in terms of *citing* papers abolishes the unsolved questions in scientometrics about the delineation of fields of science in terms of journals and normalization when comparing among different journals. Using publication and citation data of seven Korean research universities, we demonstrate the advantages and the differences in the rankings, explain the possible statistics, and suggest ways to visualize the differences in (citing) audiences in terms of a network.

---

[1] Amsterdam School of Communication Research, University of Amsterdam, Kloveniersburgwal 48, 1012 CX Amsterdam, The Netherlands; loet@leydesdorff.net; http://www.leydesdorff.net.
[2] Department of Education, Seoul National University, San 56-1, Sillim-Dong, Gwanak-Gu, Seoul 151-742, South Korea; tel: +82-2-880-7643; email: jcs6205@snu.ac.kr.

**Introduction**

In this study, for the first time we apply a new method for the fractional attribution of citations to universities as institutional units of analysis. This method was originally developed for interdisciplinary comparisons among journals, but it can be generalized to the evaluation of any inter- or multidisiciplinary grouping. The method improves previous methods for counting citations by normalizing for fields of science no longer in terms of *a priori* defined subject categories, but in terms of the sets of *citing* papers. The field of influence of the cited paper can thus be specified as the collection of its citing papers. Differences in citation behavior among the authors of citing papers are approximated by counting each reference relative to the length of the respective reference list. Thus, one can normalize the impact of publications in terms of the current state of the art at the research front.

By fractionally counting the contributions to overall citation, one can correct for the well-known problem that, for example, mathematicians provide significantly fewer references in their papers than biomedical scientists. A paper in mathematics therefore has a much lower "citation potential" (Garfield, 1979). Paradoxically, therefore, a university might improve its citation ranking by closing its mathematics departments. Similarly at the level of specialties, toxicology journals have lower impact factors than journals in immunology. Thus, research portfolios can make a difference for the ranking. For example, part of the lower ranking of universities in Asian countries is caused by differences in their research portfolios: Western countries are more oriented toward the biomedical sciences, with higher journal coverage and higher citation rates than the natural sciences and engineering (Liu *et al.*, 2004; Park *et al*., 2005).

Conceptually, citations can be considered as a measure not of intrinsic quality, but of impact. Impact has to be normalized in terms of the receiving fields. However, fields cannot be delineated clearly using scientometric methods (Leydesdorff, 2006). This new method abolishes the decades-old question of how to delineate fields of science (for the purpose of normalization) in favor of defining fields at the level of individual papers. The citing papers generate the impact and define the field of influence. This delineation has also been called source-normalization (Moed, 2010a; Small & Sweeney, 1985; Zitt & Small, 2008; Zitt, 2010). Fractional counting provides a counting rule that enables us to test differences statistically for their significance (Opthof & Leydesdorff, 2010; Spaan, 2010; Van Raan *et al*., 2010). In other words, we obtain distributions for which one can compute among other statistics a mean or a median, but also specify uncertainty (or, in other words, error in the measurement).

In this study, we apply this indicator as a first example to seven Korean universities which according to Shin (2009a) can be considered as "research universities" potentially competing at the international level. In addition to potentially changing the rankings, fractional counts of the citations can be tested against one another using an *ex post* test for multiple comparisons in ANOVA. By drawing a graph with connected nodes for the homogenous (that is, not significantly different) groupings, we provide a means for visualizing these results as a network using, for example, Pajek.[3] In addition to a ranking, our methodology thus allows for visualizing the extent to which differences in the fields in terms of the citing papers are statistically significant.

**The public-policy context**

The ranking of universities emerged during the mid-2000s in the contexts of the emerging knowledge-based economy (Foray, 2004; Foray & Lundvall, 1996; Leydesdorff, 2010) and increasing competition for world-class universities (Altbach, 2009; Halffman & Leydesdorff, 2010; Shin, 2009b). Policy makers and higher-education scholars consider world-class research universities as a source of national competitiveness. More recently, some countries (e.g., Germany, China, Korea, and Japan) began to provide special research funds to promote world-class "research centers." But what is a "world-class research university"?

The Academic Ranking of World Universities (ARWU) of the Shanghai Jiao Tong University started in 2004, the THES-QS ranking in 2005,[4] the Leiden rankings in 2008, and the rankings of Taiwan Higher Education and Accreditation Council also in 2008. This type of global ranking of universities always contains a component which focuses exclusively on research performance in terms of publications and citations. With the exception of the Leiden Ranking (which is based exclusively on these two indicators), this component is then weighted with other components (such as indicators of higher education). Unlike reputation-based rankings, however, the scientometric rankings have been less criticized by academics (e.g., Marginson, 2009) because they are based on quantifiable data. Components such as institutional reputation are difficult to operationalize and the measurement is virtually impossible to reproduce.

---

[3] Pajek is a network analysis and visualization program that is freely available for non-commercial usage at http://vlado.fmf.uni-lj.si/pub/networks/pajek/.
[4] In 2010, the two rankings of the THES and QS were uncoupled because THES decided to partner with Thomson Reuters for these rankings.

Yet, the normalization issues have remained a problem within the scientometric community to the extent that the Leiden Rankings 2010 (at http://www.socialsciences.leiden.edu/cwts/products-services/leiden-ranking-2010-cwts/#world) offer two different scales which are both based on the same bibliometric indicators (publications and citations). Although the two rankings are highly correlated at the aggregated level, differences for individual units may be large (Leydesdorff & Opthof, 2010a; Van Raan *et al*., 2010). For example, the University of Göttingen scores on the 12$^{th}$ position on the one scale of the Leiden ranking, and 125$^{th}$ on the other (Waltman *et al*., 2010). In our opinion, fractional counting offers a means to solve this problem by abolishing the need to normalize in terms of fields of science.

**The output measurement of multidisciplinary units**

Although the two prevailing output measures for research (publications and citations) seem straightforward, their normalization is beset with a number of problems. How should one, for example, attribute a paper coauthored between institutional units of analysis to each of the contributing partners: proportionally, that is, as a fraction, or by providing each unit with one whole count?

The method of fractional counting was first proposed by Price & DeBeaver (1966) for the proportionate attribution of co-authorships to papers, and has since been used more extensively in research evaluations (e.g., National Science Board, 2010; cf. Narin, 1976). Fractional counting may lead to a perspective very different from integer counting (Anderson *et al*., 1988; Leydesdorff, 1988). In summary, fractional counting would provide a negative incentive to coauthoring, whereas integer counting leads to a sum total which is larger than 100% of the set (because the same paper can be counted more than once). Nevertheless, this problem is technically surmountable (Braun *et al*., 1991).

More serious is the delineation among fields of science. Despite many vigorous attempts to cluster the aggregated journal-journal citation matrix in terms of fields of science, an unambiguous classification has hitherto failed to emerge (Leydesdorff, 2006). The journal system is "nearly decomposable" (Simon, 2003), but the overlap between otherwise discrete sets is important because it contains, among others, multidisciplinary journals such as *Science* and *Nature* (Narin *et al*., 1972). The boundaries among sets of journals are fuzzy and multi-dimensional. Forcing a classification upon the journal system leads to error (Rafols *et al*., 2010) and notably inter- or multidisciplinary work may suffer from the external perspective of one classification or another (Laudel & Orrigi, 2006).

Many evaluations are based on the so-called Subject Categories (SC) provided by the Institute of Scientific Information (ISI) of Thomson Reuters, the producer of the *Science Citation Indexes*. These subject categories for journals, however, lack an analytical base (Boyack *et al*., 2005; Pudovkin & Garfield, 2002, at p. 1113n.; Rafols & Leydesdorff, 2009) and are not literary-warranted, that is, systematically updated with reference to new literature (Bensman & Leydesdorff, 2009; Chan, 2005). Increasingly, one can use paper-by-paper indexed categories such as the Medical Subject Headings (MeSH) of the bibliographic database MedLine (Bornmann *et al*., 2008, at p. 98), but these indexes are field-specific hitherto and therefore not apt for comparisons across disciplines.

Following up on a proposal by Zitt & Small (2008), Moed (2010a) proposed ending the quest for a field definition in terms of journals by defining the system of reference for each paper as its field of impact in terms of where it is cited. Thus, if a paper in a biochemistry journal is also cited in organic chemistry journals, then this citing practice makes the paper interdisciplinary between bio- and organic chemistry more so than a paper that is cited in only one of these two fields. In the context of the noted controversy about proper normalization (Moed, 2010b; Opthof & Leydesdorff, 2010; Van Raan *et al*., 2010), Leydesdorff & Opthof (2010a and 2010b) subsequently proposed using fractional counting in terms of the reference lists in the citing papers. This solves the problem of the differences in "citation potential" among fields of science: a reference provided among 40 others counts only for $1/40^{th}$ in the overall citation, while a reference (e.g., in mathematics) provided among only six others would count for $1/6^{th}$.

Applying this normalization to the set of journals contained in the *Science Citation Index*, Leydesdorff & Bornmann (2010) have shown that fractional counting of the citations used as numerators of the impact factor reduces the in-between group variance in the impact factors among the 13 major fields of science used for the construction of the *Science and Engineering Indicators* of the US National Science Board (2010) by 81%. The remaining in-between group variance is *not* significantly different from zero. In other words, fractional counting serves the purpose of normalization among fields of science.

Impact factors by definition are based on citations to only the two preceding years (Garfield, 1972). Differences among fields of science are also caused by differences in the aging of papers (Leydesdorff, 2008). By taking all preceding years into account, that is, using the citations/publications ratio for each journal, however, the in-between group variance was *not* further reduced using the same test, but increased. Thus, the dynamic differences in citation potentials among journals (caused by differences in cited half-lives) did not statistically contribute to the differences among fields of science.

In summary, we found it urgent to introduce our new measure to the institutional evaluation using a set containing sufficient variety in terms of the disciplinary composition. In the case of Korea, we had access to a well-defined set which was used in a previous exercise (Shin, 2009a; Shin & Cummings, 2010) and to which we could add the fractional citation counts in order to see whether and how this would change the rankings. As noted, our claim is that by using this new indicator we can not only rank the universities, but also indicate whether or not differences in their impacts are significant.

The purpose of this study is to explain the methodology using a limited and straightforward case.

**Data and methods**

We harvested publication data (2005-2007) and citation data (2009) for seven Korean research universities from the *Science Citation Index-Expanded, the Social Science Citation Index* (*SSCI*), and the *Arts & Humanities Citation Index* (*A&HCI*) combined at web-interface of the ISI-*Web of Science* (*WoS*). Among these seven universities, two universities (KAIST and POSTECH) are focused on science and engineering, while the other five provide comprehensive academic programs. We used only the so-called "citable items" among the publications; that is: journal articles, letters, proceeding papers, and reviews.

At the time of data collection (April 2010),[5] 2009 was the last complete year available in terms of citations. Using the institutional addresses for the seven universities, we first collected the sets for the three years combined (2005-2007) and then used the possibility to create a "Citation Report" at the Web-interface of the *WoS* for collecting the citing documents with 2009 as the year of publication. The descriptive statistics of this data is combined with statistical information from other sources in Table 1. According to the data downloaded from the *Web of Science*, these seven universities (7,528 fte faculty members) produced a total of 42,840 papers in the period 2005-2007 which were cited 83,600 times in 2009. We will further discuss these statistics in the results section.

---

[5] At this date not all issues with publication year 2009 may have been entered into the database.

**Table 1. Research Performance and Global Ranking of Korean Universities**

| University | Number of Faculty | Publications | Citations | Citations/ Publications | Citations/ Faculty | ARWU Ranking | Leiden Ranking | THES-QS |
|---|---|---|---|---|---|---|---|---|
| POSTECH | 224 | 2,941 | 6,715 | **2.283** | 29.978 | 303-401 | - | 134 |
| SNU | 1,733 | 12,814 | 28,709 | **2.240** | 16.566 | 152-200 | 57 | 47 |
| Yonsei | 1,677 | 6,809 | 13,445 | **1.975** | 8.017 | 201-302 | 159 | 151 |
| Korea | 1,246 | 5,911 | 10,682 | **1.807** | 8.573 | 303-401 | 216 | 211 |
| KAIST | 399 | 4,776 | 8,268 | **1.731** | 20.722 | 201-302 | 199 | 69 |
| SKK | 1,118 | 5,239 | 9,063 | **1.730** | 8.106 | 303-401 | - | 357 |
| Hanyang | 1,131 | 4,350 | 6,718 | **1.544** | 5.940 | 303-401 | 245 | 339 |

Notes:
a) The faculty data is from Korean Ministry of Education and Human Resource Development in 2006.
b) Publication data are from 2005 to 2007 and citation data are the references to these papers in 2009.
c) ARWU is based on the 2010 ranking,[6] "Leiden" on 2008,[7] and Times-QS ranking is for 2009[8] because these were the latest available at the time of this research.

A dedicated routine was written to attribute the fractional counts in the citing documents to the cited universities. The fractional counts thus were classified into seven respective groupings. The citation distribution of each university is normal and the aggregated citations of the seven universities are also distributed normally. We can therefore apply an analysis of variance (ANOVA) to test how these citation distributions (in terms of fractions) differ among universities.

ANOVA enables us to test *ex post* whether differences in the distributions can also be considered significant. We will use Dunnett's C test when the variance in the distributions is not homogeneous (using Levene's test) and otherwise the equivalent Tukey test.

---

[6] Available at http://www.arwu.org/ARWU2010.jsp.
[7] The Leiden Rankings 2010 provides two rankings based on the so-called old and new crown indicators; at http://www.socialsciences.leiden.edu/cwts/products-services/leiden-ranking-2010-cwts/#world.
[8] The THES and QS rankings are different in 2010; at http://www.timeshighereducation.co.uk/world-university-rankings/ and http://www.topuniversities.com/university-rankings/world-university-rankings/2010/results, respectively.

**Table 2. Citations by Different Citation Measures**

| University | *ic* (a) | *fc* (b) | *ic/p* (c) | *100\*(fc/p)* (d) | *ic/fac* (e) | *100\*fc/fac* (f) | *p/fac* (g) |
|---|---|---|---|---|---|---|---|
| POSTECH | 6,715 | 212.90 | 2.28 | 7.24 | 29.84 | 95.04 | **13.13** |
| SNU | 28,709 | 905.52 | 2.24 | 7.07 | 15.98 | 52.25 | 7.39 |
| Yonsei | 13,445 | 439.74 | 1.97 | 6.46 | 8.46 | 26.22 | 4.06 |
| Korea | 10,682 | 345.71 | 1.81 | **5.85** | 9.06 | 27.75 | 4.74 |
| KAIST | 8,268 | 289.01 | 1.73 | **6.05** | 20.72 | 72.43 | **11.97** |
| SKK | 9,063 | 301.17 | 1.73 | 5.75 | 8.51 | 26.94 | 4.69 |
| Hanyang | 6,718 | 234.67 | 1.54 | 5.39 | 5.85 | 20.75 | 3.85 |

*ic* = integer-counted number of citations; *fc* = fractionally counted number of citations; *p* = number of publications; *fac* = fte of faculty.

Table 2 provides descriptive statistics of the fractionally counted versus integer counted citations for the seven universities under study. In the case of these seven universities, the rank-order based on total numbers of citations is not affected by choosing either integer or fractional counting (columns *a* and *b*). However, there is an effect on the impact parameter of fractional counting per publication in column *d*. Note also the irregular (and highly correlated) behavior of the productivity parameters in columns *e* to *g*. Let us now proceed to discuss these results in more detail.

**Results**

As can perhaps be expected, the total numbers of faculty, publications, and citations in Table 1 are both highly and significantly correlated ($\rho > 0.89$; $p < 0.01$). These three measures indicate *size*. However, the impact measure citations/publications is *not* significantly correlated to any of these size indicators. The numbers of publications/faculty or citations/faculty are *negatively* correlated with size parameters and not with the impact indicators.

In summary, one can distinguish three independent dimensions: (*i*) *size*—indicated as the total number of publications ($\Sigma p$), citations ($\Sigma c$) or faculty (*fac*); (*ii*) *impact* or citation per publication (*c/p*); and (*iii*) productivity, that is publications or citations per faculty (*p/fac* or *c/fac*, respectively). Size and impact are not correlated. Size and productivity are negatively correlated: the smaller universities are relatively more productive than the larger ones.

Among the three global rankings included in Table 1, only the Leiden Rankings 2008 correlates significantly ($\rho = 0.90$; $p < 0.05$) with our rankings in terms of publications, citations, and citations/publications.[9] Size (in terms of faculty) is again not correlated with any of these rankings. With the exception of the Leiden Rankings, the various rankings are based on weighting and normalization schemes among various indicators which make it impossible for an outsider to reproduce them.

Fractional counting produces the same rankings as integer counting when the citations are aggregated (columns *a* and *b* in Table 2). However, ranking by the two counting methods generates different results when the numbers of citations are divided by the number of papers (columns *c* and *d*). In this case, KAIST scores higher than the Korea University. KAIST then also ranks indisputably higher than SKK to which it was tied in the ranking based on integer counting. Since KAIST is a small and technologically oriented (i.e., specialized) university, the normalization for the portfolio thus matters.

---

[9] When fractional citations are counted, the rank-order correlation with the Leiden Rankings is unity.

**Table 3. Spearman Rank Correlations among Citation Measures**

|       | p (a)      | ic (b)     | fc (c)     | ic/p (d) | fc/p (e) | c/fac (f) |
|-------|------------|------------|------------|----------|----------|-----------|
| *fac* | .893(**)   | .893(**)   | .893(**)   | .126     | .000     | -.464     |
| *p*   |            | 1.000(**)  | 1.000(**)  | .234     | .143     | -.214     |
| *ic*  |            |            | 1.000(**)  | .234     | .143     | -.214     |
| *fc*  |            |            |            | .234     | .143     | -.214     |
| *ic/p*|            |            |            |          | .937(**) | .595      |
| *fc/p*|            |            |            |          |          | .714      |

** Correlation is significant at the 0.01 level (2-tailed).
* Correlation is significant at the 0.05 level (2-tailed).

Integer and fractional counting of the citations per faculty generate the same rankings (column *f* in Table 3). The correlations with all size parameters are negative. In terms of these measures, KAIST and POSTECH, which are both *small* (in size) and *science and technology focused* (in programs), perform better than larger and comprehensive universities (column *g* in Table 2). The rankings between citations/faculty and publications/faculty are precisely similar ($\rho = 1.0$). Thus, the different citation rates may find their origin in the different publication rates (per faculty) of these smaller universities.

This conclusion is important because the normalization across fields by using fractional citation counts does not normalize for differences in *publication* behavior among authors in different fields of science. Faculty in the social sciences and humanities tend to publish with fewer co-authors, whereas faculty in engineering and natural sciences publish more papers with multiple co-authors (Park *et al*., 2010). Considering the publication patterns across disciplines, faculty in KAIST and POSTECH seem to have more opportunities for publishing. However, the size and productivity aspects are analytically independent from the impact: when normalized for field differences (by using fractional counting), the rank-order correlation ($\rho$) between the *fc/p* parameter and the number of faculty happens to be precisely 0.000.

**Differences of Citation Impacts across Universities**

In addition to correcting for field differences, the fractional citation counts provide us with distributions of values which we averaged above, but which contain further information that can be considered in the evaluation. Are universities—in this case, our units of analysis—significantly different in terms of their citation impact? Using the non-parametric Kruskall-Wallis test among these seven groups taught us that the differences are significant ($p = 0.000$). Using a so-called *post hoc* test in ANOVA enables us to specify precisely which universities differ significantly from others in terms of these fractionated citation patterns.[10] The results are provided in Table 4. We highlighted the third column, which indicates whether or not the means of the distributions are significantly different from each other.

---

[10] The fractional citation distribution of each individual university is normal, and the total citation of the seven universities is also a normal distribution.

**Table 4: Multiple Comparisons of the Seven Korean Universities**

| (I) University | (J) University | Mean Difference (I-J) | Std. Error | 95% Confidence Interval Lower Bound | 95% Confidence Interval Upper Bound |
|---|---|---|---|---|---|
| SNU | KAIST | **-.003413690*** | .000389314 | -.00456175 | -.00226563 |
|  | POSTECH | **-.000162739** | .000324494 | -.00111968 | .00079420 |
|  | Yonsei | **-.001164976*** | .000327229 | -.00212988 | -.00020008 |
|  | Korea | **-.000821823** | .000334477 | -.00180813 | .00016448 |
|  | Hanyang | **-.003389721*** | .000398377 | -.00456457 | -.00221487 |
|  | SKK | **-.001689377*** | .000340328 | -.00269296 | -.00068580 |
| KAIST | SNU | **.003413690*** | .000389314 | .00226563 | .00456175 |
|  | POSTECH | **.003250951*** | .000441871 | .00194782 | .00455408 |
|  | Yonsei | **.002248714*** | .000443883 | .00093973 | .00355770 |
|  | Korea | **.002591867*** | .000449254 | .00126703 | .00391671 |
|  | Hanyang | **.000023969** | .000498656 | -.00144664 | .00149457 |
|  | SKK | **.001724313*** | .000453626 | .00038656 | .00306207 |
| POSTECH | SNU | **.000162739** | .000324494 | -.00079420 | .00111968 |
|  | KAIST | **-.003250951*** | .000441871 | -.00455408 | -.00194782 |
|  | Yonsei | **-.001002236** | .000388283 | -.00214728 | .00014281 |
|  | Korea | **-.000659084** | .000394411 | -.00182222 | .00050405 |
|  | Hanyang | **-.003226982*** | .000449877 | -.00455377 | -.00190020 |
|  | SKK | **-.001526638*** | .000399385 | -.00270446 | -.00034881 |
| Yonsei | SNU | **.001164976*** | .000327229 | .00020008 | .00212988 |
|  | KAIST | **-.002248714*** | .000443883 | -.00355770 | -.00093973 |
|  | POSTECH | **.001002236** | .000388283 | -.00014281 | .00214728 |
|  | Korea | **.000343152** | .000396664 | -.00082654 | .00151285 |
|  | Hanyang | **-.002224745*** | .000451853 | -.00355728 | -.00089221 |
|  | SKK | **-.000524401** | .000401610 | -.00170870 | .00065990 |
| Korea | SNU | **.000821823** | .000334477 | -.00016448 | .00180813 |
|  | KAIST | **-.002591867*** | .000449254 | -.00391671 | -.00126703 |
|  | POSTECH | **.000659084** | .000394411 | -.00050405 | .00182222 |
|  | Yonsei | **-.000343152** | .000396664 | -.00151285 | .00082654 |
|  | Hanyang | **-.002567898*** | .000457130 | -.00391601 | -.00121978 |
|  | SKK | **-.000867554** | .000407537 | -.00206935 | .00033425 |
| Hanyang | SNU | **.003389721*** | .000398377 | .00221487 | .00456457 |
|  | KAIST | **-.000023969** | .000498656 | -.00149457 | .00144664 |
|  | POSTECH | **.003226982*** | .000449877 | .00190020 | .00455377 |
|  | Yonsei | **.002224745*** | .000451853 | .00089221 | .00355728 |
|  | Korea | **.002567898*** | .000457130 | .00121978 | .00391601 |
|  | SKK | **.001700344*** | .000461428 | .00033954 | .00306115 |
| SKK | SNU | **.001689377*** | .000340328 | .00068580 | .00269296 |
|  | KAIST | **-.001724313*** | .000453626 | -.00306207 | -.00038656 |
|  | POSTECH | **.001526638*** | .000399385 | .00034881 | .00270446 |
|  | Yonsei | **.000524401** | .000401610 | -.00065990 | .00170870 |
|  | Korea | **.000867554** | .000407537 | -.00033425 | .00206935 |
|  | Hanyang | **-.001700344*** | .000461428 | -.00306115 | -.00033954 |

*. The mean difference is significant at the 0.05 level.

These results can be summarized graphically using Figure 1. In this graph, universities which are connected are significantly similar in terms of their fractional citations by relevant audiences. As the figure shows, five universities are linked directly or indirectly, but the other two universities are linked to each other separately from these five. The figure implies that the five universities have similarities in terms of their citation patterns, and the other two universities have similarities with each other.

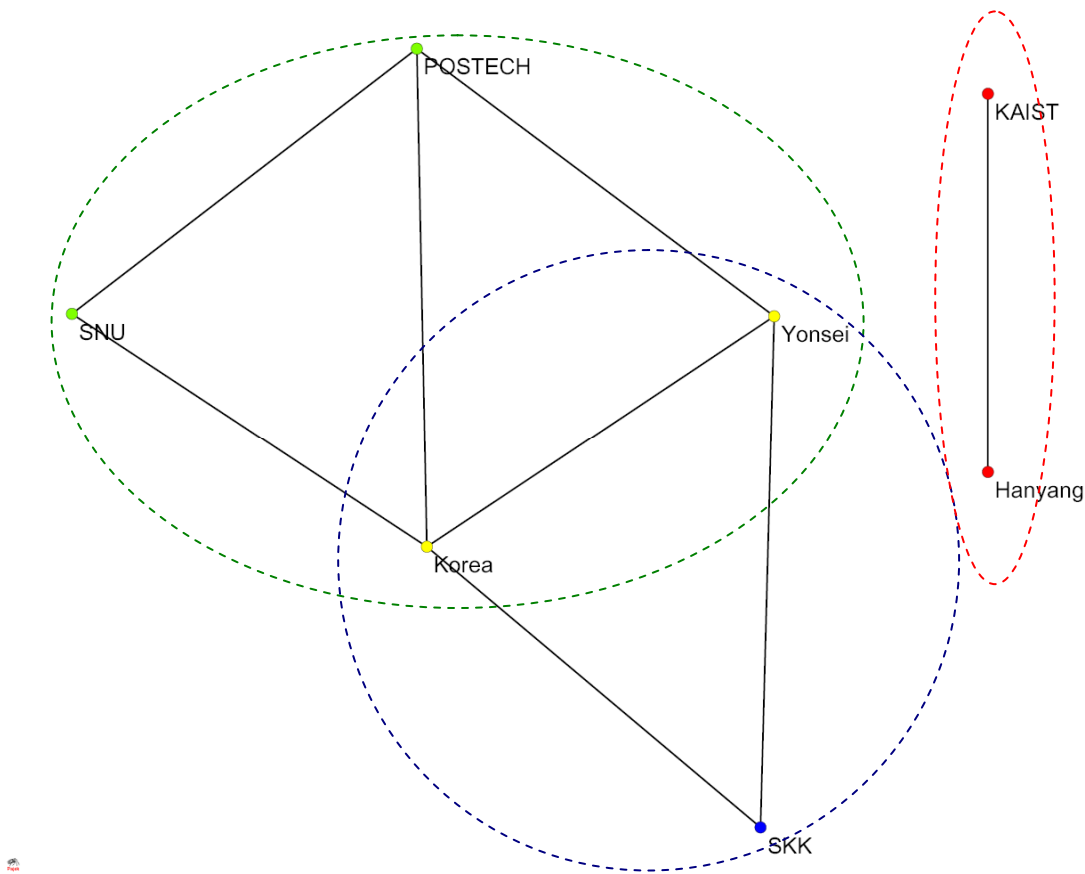

**Figure 1**: Universities which are not significantly different in their impact are linked in this graph. Homogenous groups are circled based on the Tukey-test.[11]

Using the Tukey test, we can also distinguish homogenous *groups* among these seven universities in terms of their citing audiences. These are indicated in Figure 1 as circles and ellipses. The interpretation is as follows: SNU is not significantly different from POSTECH and Korea in terms of its fractional being-cited patterns, but significantly different from the four other universities, including Yonsei and SKK. Hanyang and KAIST have a citation impact significantly different from the other five universities. Yonsei and the Korea University can be considered as belonging to two groups in terms of their relevant audiences.

---

[11] The results of the Tukey test were identical to the ones of Dunnett's C-test and can therefore be used.

Note that we test in this case the properties of citing documents. (The cited documents are used only for delineating the sample.) These distributions can be considered as proxies for the citation behavior of relevant audiences (Zitt, 2010; Zitt & Small, 2008). The results teach us that these audiences are not homogenous in terms of their citing behavior, and thus fractional counting of their citations is further legitimated. Using integer counting, we would thus be comparing apples with oranges because the differences among the citation behavior in these sets are statistically significant.

The homogeneous grouping together of KAIST and Hanyang as different from the other five can perhaps be explained institutionally in terms of the ratio of engineering versus natural sciences at these two universities. These ratios are 70:30 for KAIST and 66:34 for Hanyang while the ratios are lower for other universities: 61:39 for POSTECH, 59:41 for SKK, 47:53 for Korea, and 44:56 for SNU and Yonsei, respectively. The participation, contribution, and positioning of engineering may thus provide a distinguishing feature among Korean universities in terms of their citation impact. However, the composition of the research portfolios is more complex than only this distinction. SNU and Yonsei, which are tied in terms of this ratio (at 44:56), are not indicated as significantly similar in this respect.

**Conclusions and discussion**

Fractional counting of the citations enabled us to rank the seven research universities of Korea in terms of their impact normalized for research portfolios, and to show that these portfolios are significantly different in terms of their disciplinary audiences. The exercise made clear that impact cannot be expected to change by changing the portfolio because the productivity, size, and impact dimensions are analytically independent. Productivity was negatively correlated to size: by hiring more engineering staff the larger universities in Korea can perhaps increase their numbers of publications and citations, but not their impact and productivity. The differences in productivity between large and small (specialized) universities in Korea are approximately a factor of two.

The position of only one of the two small and specialized universities, KAIST, had to be changed in the ranking when using fractional instead of integer counts of the citations. This difference cannot be explained only in terms of the differences in the citation potentials among these universities. The average length of the reference list (= *ic*/*fc*) is equal to 28.6 for both KAIST and Hanyang, but above 30 for the other universities. Thus, one can expect that these two universities would be undervalued using integer counting. However, this field effect is only the case for KAIST. Hanyang, in other words, remains at the bottom of the hierarchy among these seven universities using both integer and fractional counting. The other specialist university with a high productivity (POSTECH) leads using either of these two counting methods, but the distance to the second in line is enhanced by using fractional counting.

We have meant this contribution mainly to show how it is possible to test and correct for in-between field differences when using citations or citations/publication as an evaluative criterion in the case of intellectually non-homogenous groups. This issue was raised in the context of a controversy about proper normalization using journals as a framework for the classification. Journals, however, are mixed bags: they may contain articles, reviews, and letters to different extents, and they are not likely to be classified unambiguously because of intellectual organization in a variety of dimensions. This can be handled analytically in a vector space (using multivariate analysis such as MDS or factor analysis), but classification necessarily sacrifices the richness of the multiple dimensions.

Classification at the level of documents such as, for example, in the Medical Subject Headings of the *MedLine* or the classifications of *Chemical Abstracts* are discipline-specific and therefore not suited for inter- or multidisciplinary comparisons. Differences in citation potential, however, are generated by differences in the citation behavior of authors in individual (or coauthored) papers. Fractional counting provides direct access to these differences in behavior and allows accordingly for the normalization.

Two problems remain. First, the differences in aging among papers may be significantly affected by disciplinary delineations and in terms of different document types (e.g., letters or reviews). Mathematics papers, for example, cite not only less, but also older literature than bio-medical papers. The organization of laboratories at a research front can be expected as a major source of differences. However, we could show that these are field-characteristic patterns of behavior which are corrected by fractional counting at the research front (Leydesdorff & Bornmann, 2010). Moed (2010c) objected that papers with zero citations would thus not be taken into account, but these non-citations can, in our opinion, not be considered as included in the audience of a paper in terms of scientometric measurement.

In some fields of science, it is considered an honor to be cited by a prestigious review which itself includes hundreds of citations. Fractional counting would not justify this effect because the contribution to overall citation would be marginalized. From this perspective, a further refinement could be to distinguish among references in reviews and articles or conference proceedings. Elsevier's SNIP indicator operates already without taking letters into account (Moed, 2010a). For the time being, however, the organization of this database in terms of document types is not sufficiently precise (Leydesdorff & Opthof, 2010c). Using the *Web-of-Science*, this next step can be made in a more comprehensive evaluation. However, this seems an industrial task more than an academic one.

**Acknowledgment**

Jung C. Shin acknowledges funding from the National Research Foundation of Korea (project Social Science Korea).